The thermal emission for knot A of radio galaxy M87 by *Chandra*

S.Osone

Chiba, Japan
osonesatoko@gmail.com

Abstract
The jet may compress an interstellar medium and soft X ray outside the jet may be absorbed by a compressed interstellar medium and thermal emission heated by a shock is expected inside the jet. Thermal emission was found for HST-1 and knot A of M87 in X ray energy spectra analysis by *Chandra* (Osone, 2017). The X ray image of the surrounding of the jet of M87 was studied by *Chandra* (Osone 2024). I find a soft X ray dip in a south region outside knot A. Therefore, by taking only a north region outside knot A as a background, where there is no soft X ray dip, I analyze X ray energy spectra for knot A by *Chandra* and confirm thermal emission with a high significance.

1.Introduction
M87 is a radio galaxy and located in the center of Virgo cluster. The distance is 16.8 Mpc (Blakeslee et al. 2009). M87 is famous for Blackhole shadow observation by EHT (EHT 2019).

M87 is observed from a radio to TeV gamma ray. M87 have an inclined jet with a length of 20”. The angular resolution is micro second in a radio, 0”.7 in an optical band and 0”.5 in an X ray by *Chandra*, 5’.2 in GeV gamma ray by *Fermi* , 2’ in TeV gamma ray by air Cherenkov detector as CTA, HESS, LHASSO, MAGIC and VERITAS, 0.1 degree in TeV gamma ray by water Cherenkov detector as HWAC and LHASSO, and 0.5 degree in TeV gamma ray by air shower arrays as LHASSO and Tibet AS $\gamma$ . Therefore, the jet cannot be resolved in high energy gamma ray. However, the origin of high energy emission can be limited by a timescale of flux variability and by simultaneous observation with multiwave length. A time scale of flux variability is 2d in TeV gamma ray (Aharonian et al. 2006). This means that the size of emitting area is $\delta ct_{val} = 3.1\text{x}10^{16} (\delta/6)$ cm. Here, c is a speed of light and δ is a beaming factor. Therefore, the origin of TeV gamma ray is considered as nucleus. TeV flare occurred in 2018 and 2022 since 2010. Algaba et al. (2024) and Cao et al.(2024) also showed nucleus as the origin of TeV gamma ray from flux variability in TeV gamma ray. A flux correlation between nucleus in radio band and TeV gamma ray in 2008 showed nucleus as origin of TeV gamma ray (Acciari et al. 2009). A flux correlation between nucleus in X ray and TeV gamma ray in 2008 and 2010, between HST-1 in X ray and TeV gamma ray in 2005 were reported (Abramowaski et al. 2012). Therefore, HST-1 is also considered as the origin of TeV gamma ray.

The standard interpretation of multiwave length energy spectra for the M87 jet is synchrotron inverse compton model of accelerated electrons. The energy spectra is described by synchrotron emission in a low energy and inverse compton model in a high energy. The identical origin of electrons expects same flux variability in all energy band. Benkhali et al. (2019) analyzed *Fermi* data from 2008 to 2016 and showed a chance probability of non flux variability is 0.018 below 10 GeV and 0.23 above 10 GeV. Cao et al.(2024) analyzed *Fermi* data from 2009 to 2024 and showed non flux variability with three months bin, but showed flux variability with a short time scale bin of day. Algaba et al.(2024) also analyzed *Fermi* data in 2018 and reported flux variability with bin of few days. However, count per short time scale is poor statistics and significance of flux variability is not reported. Flux variability in GeV gamma ray is debated yet. There are some models which show non flux variability in GeV gamma ray. One is a pion decay as an interaction between accelerated protons and an interstellar medium. This energy spectra peaks at 80 MeV, which is a half energy of pion. However, energy spectra of M87 have not been reported below 1 GeV by *Fermi*. Another is non thermal bremsstrahlung as an interaction between accelerated electrons and an interstellar medium. Osone (2017) showed no contribution of non thermal bremsstrahlung from both nucleus and HST-1 to observed flux with *Fermi*.

Dainotti et al. (2012) made a merged image with archival data from 2000 to 2009 by *Chandra* and reported that there is a soft X ray dip outside the jet between knot E and knot F as the interaction between the jet and an interstellar medium. There is a pile up problem for CCD of *Chandra*. A pile up distorts both an energy spectra and an image. For a merged image, they removed data with a visible pile up line originated from nucleus or HST-1. However, they did not remove data with pile up event in a count rate. For correct image analysis, pile up event have to be removed completely. There are two kinds of apprant movements. One is a jet apprant velocity, the other is a precession. When an image is merged, these apprant movements have to be considered. Osone(2024) used archive data from 2000 to 2018 with *Chandra* and made a merged image by a removal of pile up event completely and a consideration of an apparent movement and confirmed soft X ray dip outside the jet between knot E and knot F.

As an interaction between a jet and an interstellar medium, thermal emission added to synchrotron emission is expected. The X-ray energy spectra of the M87 jet has been analyzed with *Chandra* (Wilson & Yang 2002; Marshall et al. 2002; Perlman & Wilson 2005;

Sun et al. 2018). They fitted energy spectra with a power law model as synchrotron emission and obtained an acceptable fit. There are two problems for their analysis, background region and statistics. Hot gas of the Virgo cluster centered at M87 has been reported with *XMM* (Belsole et al. 2001). The background for the jet is the sum of Cosmic X-ray background, galactic emission, solar activity, non X-ray background and hot gas of Virgo cluster. The emission of hot gas of Virgo cluster depends on a distance from the center (Bohringer et al. 2001). Therefore, background should be taken according to a distance from the nucleus. There are two problems with poor statistics. Any models tend to be acceptable with poor statistics. The other is contamination of hot gas with poor statistics. There is some contamination of hot gas of the cluster shown as a line feature on the energy spectra which subtract background for obsID 1808. The subtraction of background is not enough with an exposure time of 13 ks. With more statistics, this contamination is expected to be less. This is the reason why high statistics by using plenty archive is needed. In Osone (2017), archival data with an exposure time of about 800 ks were used and a correct subtraction of background was done. Osone (2017) found thermal emission added to synchrotron emission for knot HST-1 and knot A.

## 2.Spectral analysis

### 2.1 soft X ray dip

Background for the jet of M87 is cosmic X ray Background, detector background, galactic emission, solar activity and hot gas of Virgo cluster. X ray emission of hot gas of Virgo cluster depends on the distance from the nucleus of M87(Bohringer et al. 2001). For energy spectra analysis of the jet of M87 in Osone (2017), background is taken in two neighbor regions of a north and a south and with almost same distance from nucleus, for each bright knots (nucleus, HST-1, knot D and knot A).

I show the merged image of the M87 jet with an exposure time of 224 ks by *Chandra* (Osone 2024) in figure 1. The two white ellipses show soft X ray dip area in a north which was found in Dainotti et al. (2012) and was confirmed in Osone (2024), and a reference region in a south, outside the jet between knot E and knot F. The two white circles show background regions for energy spectra analysis of knot A in Osone (2017). Here, I find the significance of soft X ray dip in a south region is 5.6 sigma against a north region, outside knot A with an exposure time of 224 ks. The process of data analysis is same with Osone (2024). The soft Xray dip in a background region has some effect on the energy spectra. When soft X ray dip exists in background, thermal emission may be appeared in the energy spectra which subtract background. Therefore, I take only a north region as background for knot A in order to confirm thermal emission. For HST-1, there is no soft

X ray dip in a background region.

A position of a source region for obsID 18232 and obsID 5826 is shown in table 1. The image of obsID18232 made by ds9 tool is shown in figure 2. The exposure time of obsID 18232 is 18 ks. The region file which shows a position of a source and its radius and that which show a position of background and its radius are made by ds9 tool. The radius of circle as source region and a background region is 1 arc second. A position of each knot is changed from data to data by apparent velocity and precession. A position of an extracted region is decided by image for each data set. A radius of an extracted region is same by a data set.

## 2.2 data

The detector used is CCD. The image is taken from 2 dimensions array of CCD and an X-ray energy spectra is taken from a deposit energy in CCD. I use archival data from 2000 to 2018. The knot A are saturated in a frame time of 3.2 sec. Therefore, a frame time of 0.4 sec is used. Data is same with Osone (2017). Data observed from 2000 Jul. to 2014 Dec. is called as a former data set and data observed from 2015 Mar. to 2018 Apr. is called as a latter data set. CIAO software of 4.15 and caldb 4.10.2 are used.

## 2.3 pile up treatment

Nucleus and HST-1 are sometimes piled up heavily in a former data set. Heavy pile up distorts X-ray energy spectra. I select data with following two steps. At first, I check an image by ds9 tool. The apparent pile up line is sometimes originated from nucleus or HST-1. The pile up line is sometimes overlapped with a source region or a background region for knot A. these data set is not used for analysis. Next, I check a pile up in count rate by a pileup_map command. This command makes a pile up information from an event data and the count/frame is calculated for a given region file by a dmstat command. The count/frame = 0.1, 0.2 corresponds to 5%, 10% pile up fraction, respectively. I select data with 5% pile up fraction. For all data of knot A, pile up fraction is less than a 5%. The used observation log for knot A is shown in table 2.

## 2.4 energy spectra

An energy spectra of source region and background region are made by a specextract command from fits data and a region file. The effective area (arf) and energy response(rmf) are also made by a specextract command from fits data and a region file. The arf is made with no weight of a count rate and a correction by PSF which are suitable conditions for a point source analysis. Arf and rmf are made for only source region. Arf and rmf are multiplied as rsp by a marfrmf command for each data set and rsp for summed data set is weighted with an exposure time by an addrmf command. Data points in energy spectra of a source are binned so that minimun counts per a bin are above 15 by a grppha command.

The summed energy spectra which subtract background is described with no line feature and a contamination of hot gas of cluster is excluded successfully.

2.5 model fitting

CCD has a sensitivity from 0.2 keV to 10 keV. There is a quantum efficiency degradation by a contamination of an optical filter. An energy range above 0.3 keV, especially above 0.5 keV is well calibrated. Therefore, a lower limit of an energy in energy spectra is set as 0.5 keV.

XSPEC tool is used for an energy spectra analysis. At first, a power law model as synchrotron emission of accelerated electrons is used. Here, an absorption in soft X-ray is caused by a photo electric effect of a neutral material in a line of a sight from us to M87. The column density by a 21 cm radio observation is $1.27 x 10^{-20} cm^{-2}$ (Bekhti et al 2016). An absorption is applied to a power law. A phabs model is used as an absorption. At first, a column density is set free. When a column density is lower than the 21 cm value, a column density is fixed to the 21 cm value and energy spectra is fitted with a power law. When a power law model is not reasonable, an APEC model as thermal bremsstrahlung with a metal is added to a power law model and energy spectra is fitted. Here, an absorption is also applied to an APEC model.

2.6 fitting result

Fitting results with a power law and a combination model of a power law and an APEC are shown in table 3 and table 4, respectively.

2.6.1 former data set

The energy spectra cannot be described with a power law model. There is a soft excess. $X^2$=465.50 for d.o.f =295 for a power law model show a chance probability of $8.7 x 10^{-10}$. A power law model is rejected statistically. Therefore, when an APEC model is added to a power law model, a reasonable fitting is obtained. The temperature of 0.2 keV and a metal abundance of 0.00 are obtained. In an APEC model, a relative abundance is fixed to a value of Grevesse & Anders (1989). A relative abundance is cannot be free because of large uncertainty.

2.6.2 latter data set

The energy spectra is well described with a power law model.

2.6.3 all data set

The energy spectra cannot be described with a power law model. There is a soft excess as shown in figure 3 (top). $X^2$=652.58 for d.o.f=355 for a power law model show the chance probability of $7.2 \times 10^{-20}$. A power law model is rejected statistically. When an APEC model is added to a power law model, a reasonable result is obtained as shown in figure 3(bottom). The temperature of 0.2 keV and a metal abundance of 0.00 are obtained. The flux ratio of an APEC model to total is 19%. In an APEC model, a relative abundance is fixed to a value of Grevesse & Anders (1989). A relative abundance is cannot be free because of large uncertainty. The metal abundance is low compared with hot gas of Virgo cluster which metal abundance is 1 solar (Belsole et al 2001). When a metal abundance is fixed to 0.1 solar or 1 solar, these models cannot be rejected statistically.

2.7 plasma density

A normalization of an APEC model is given as $10^{-14} \times n_e n_i V / 4\pi D_A{}^2(1+z)^2$. Here, $n_e$ is an electron density in units of $cm^{-3}$ and $n_i$ is an ion density in units of $cm^{-3}$. $V$ is a volume of the extracted region in units of $cm^3$. $D_A$ is an angular diameter distance to M87 in units of cm. $z$ is a redshift. An electron density $n_e$ is assumed to be equal to an ion density $n_i$ and an ion density $n_i$ is derived as 12.6(+2.2, -2.6) $cm^{-3}$ for former data set, 11.4(+1.6, -1.4) $cm^{-3}$ for all data set. When a metal abundance is fixed to 0.1 and 1 solar for all data set, a plasma density is 4.7(+0.2, -0.3)$cm^{-3}$ and 1.6(±0.1)$cm^{-3}$, respectively. The error is 90% confidence level statistical error.

3. Conclusition

Osone (2017) found thermal emission added to synchrotron emission for HST-1 and knot A in an X ray energy spectra by *Chandra*. In X ray merged image of the M87 jet with an exposure time of 224 ks by *Chandra* (Osone 2024), I find a soft Xray dip in a south region outside knot A. Therefore, for X ray energy spectra analysis, I take only a north region as background, where there is no soft Xray dip and make an X ray energy spectra for knot A with same data with Osone (2017). I confirm thermal emission for knot A.

Acknowledgement

I thank *Chandra* archival data center and *Chandra* software team and CXC Help desk for a kind support.

Table 1 The position of nucleus, HST-1, knot D and knot A in obsID 18232 and knot E and knot F in obsID 5826. 1" corresponds to 78 pc

| | (RA, DEC)(J2000) | distance from nucleus |
|---|---|---|
| nucleus | ($12^h30^m49^s.46$, 12°23'28".0) | |
| HST-1 | ($12^h30^m49^s.38$, 12°23'28".4) | 1".2(94 pc) |
| D | ($12^h30^m49^s.27$, 12°23'28".9) | 2".9(226 pc) |
| E | ($12^h30^m49^s.03$, 12°23'30".2) | 6".8(530 pc) |
| F | ($12^h30^m48^s.87$, 12°23'31".0) | 9".4(733 pc) |
| A | ($12^h30^m48^s.65$, 12°23'32".3) | 12".8(998pc) |

Table 2 The observation log used for spectra analysis for knot A. Data with a frame time of 0.4 sec is used.

| obsID | PI | obs date | number |
|---|---|---|---|
| former data set | | | |
| 1808 | Wilson | 2000.7 | 1 |
| 3084~3088 | Harris | 2002.2~2002.7 | 5 |
| 3975~3982 | Harris | 2002.11~2003.8 | 8 |
| 4917~4919 | Birreta | 2003.11~2004.2 | 3 |
| 5740 | Birreta | 2005.4 | 1 |
| 6301~6303,6305 | Birreta | 2006.2~2006.8 | 4 |
| 7351 7352 | Birreta | 2007.3~2007.5 | 2 |
| 8510~8517 | Harris | 2007.2~2007.3 | 8 |
| 8575~8581 | Birreta | 2007.11~2008.8 | 7 |
| 10282~10288 | Harris | 2008.11~2009.12 | 7 |
| 11512~11520 | Harris | 2010.4~2010.5 | 9 |
| 13964~13965 | Harris | 2011.12~2012.2 | 2 |
| 14973~14974 | | 2012.12~2013.3 | 2 |
| 16042~16043 | | 2013.12~2014.4 | 2 |
| 17056~17057 | | 2014.12~2015.3 | 2 |
| exposure time | 304ks | | |
| latter data set | | | |
| 18809~18813 | Cheng | 2016.3 | 5 |
| 18232~18233 | Russell | 2016.2~2016.4 | 2 |
| 18781~18783 | | 2016.2~2016.4 | 3 |
| 18836~18838 | | 2016.4 | 3 |
| 18856 | | 2016.6 | 1 |
| 20034~20035 | Neilsen | 2017.4 | 2 |
| 19457~19458 | Wong | 2017.2 | 2 |
| 21075~21076 | | 2018.4 | 2 |
| 20488~20489 | Cheng | 2018.1~2018.3 | 2 |
| exposure time | 385ks | | |
| total exposure time | 689ks | | |

Table 3 Fitting parameters with a power law for summed energy spectra. An error is 90% confidence level statistical error.

| | $N_H$ | photon index | 1keV flux | $\chi^2$/d.of |
|---|---|---|---|---|
| | ($\times10^{20}$ cm$^{-2}$) | | (ph/s/cm$^2$/keV) | (d.o.f) |
| former data set | 0.00 | $2.43^{+0.02}_{-0.02}$ | $2.35^{+0.02}_{-0.02}$($\times10^{-4}$) | 1.370(294) |
| | 1.27 fix | $2.48^{+0.02}_{-0.01}$ | $2.44^{+0.03}_{-0.02}$($\times10^{-4}$) | 1.578(295) |
| latter data set | 0.00 | $2.38^{+0.02}_{-0.02}$ | $1.95^{+0.02}_{-0.03}$($\times10^{-4}$) | 1.095(298) |
| | 1.27 fix | $2.42^{+0.02}_{-0.03}$ | $2.02^{+0.03}_{-0.03}$($\times10^{-4}$) | 1.171(299) |
| all data | 0.00 | $2.44^{+0.01}_{-0.02}$ | $2.20^{+0.02}_{-0.02}$($\times10^{-4}$) | 1.545(354) |
| | 1.27 fix | $2.48^{+0.02}_{-0.01}$ | $2.29^{+0.01}_{-0.02}$($\times10^{-4}$) | 1.838(355) |

Table 4 Fitting parameters with a combination model of a power law and an APEC for summed energy spectra of knot A. A column density of an absorption is fixed to the 21cm observation value. An error is 90% confidence level statistical error.

| | photon index | 1keV flux | $kT$ | abundance | normalization | $\chi^2$/d.o.f |
|---|---|---|---|---|---|---|
| | | (ph/s/cm$^2$/keV) | (keV) | (solar) | | (d.o.f) |
| former data set | $2.28^{+0.04}_{-0.04}$ | $2.13^{+0.06}_{-0.04}$($\times10^{-4}$) | $0.19^{+0.03}_{-0.03}$ | 0.00 | $2.88^{+1.18}_{-0.94}$($\times10^{-3}$) | 0.965(292) |
| all data | $2.28^{+0.03}_{-0.02}$ | $2.00^{+0.02}_{-0.08}$($\times10^{-4}$) | $0.21^{+0.00}_{-0.02}$ | $0.00^{+0.01}$ | $2.50^{+0.74}_{-0.62}$($\times10^{-3}$) | 0.919(352) |
| | $2.32^{+0.02}_{-0.02}$ | $2.05^{+0.03}_{-0.03}$($\times10^{-4}$) | $0.20^{+0.01}_{-0.01}$ | 0.1 fix | $4.13^{+0.41}_{-0.40}$($\times10^{-4}$) | 1.011(353) |
| | $2.33^{+0.02}_{-0.02}$ | $2.07^{+0.03}_{-0.03}$($\times10^{-4}$) | $0.20^{+0.01}_{-0.01}$ | 1.0 fix | $4.76^{+0.46}_{-0.46}$($\times10^{-5}$) | 1.044(353) |

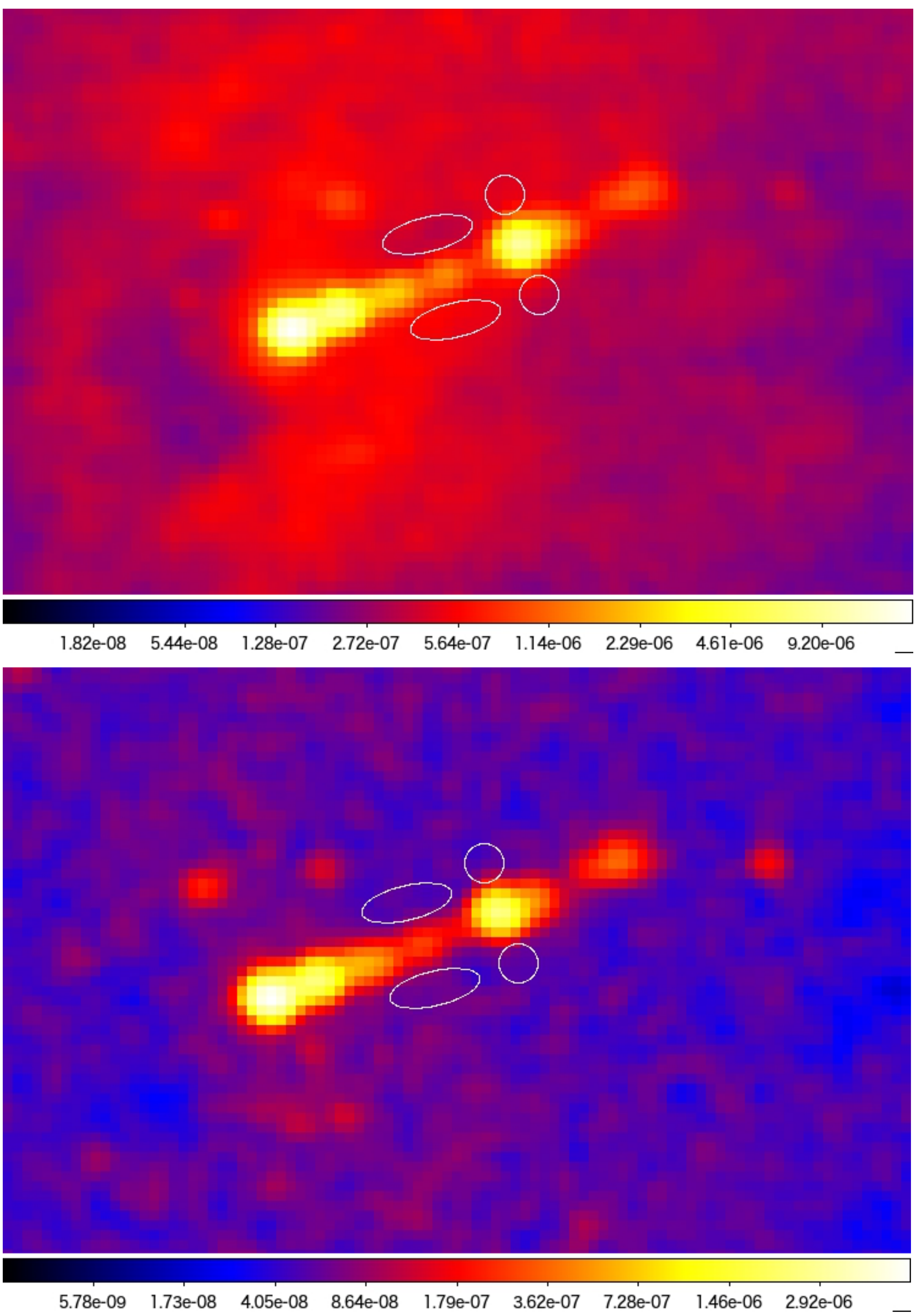

Figure 1 The merged image for the jet of M87(Osone 2024). The energy range is 0.5-2 keV(top) and 2-7 keV(bottom). The two ellipses are a soft X ray dip region in a north which was found in Dainotti et al. (2012) and was confirmed in Osone (2024), and a reference region in a south outside the jet between knot E and knot F. The two white circles are background regions for knot A in Osone (2017). There is a soft X ray dip in a south region outside knot A.

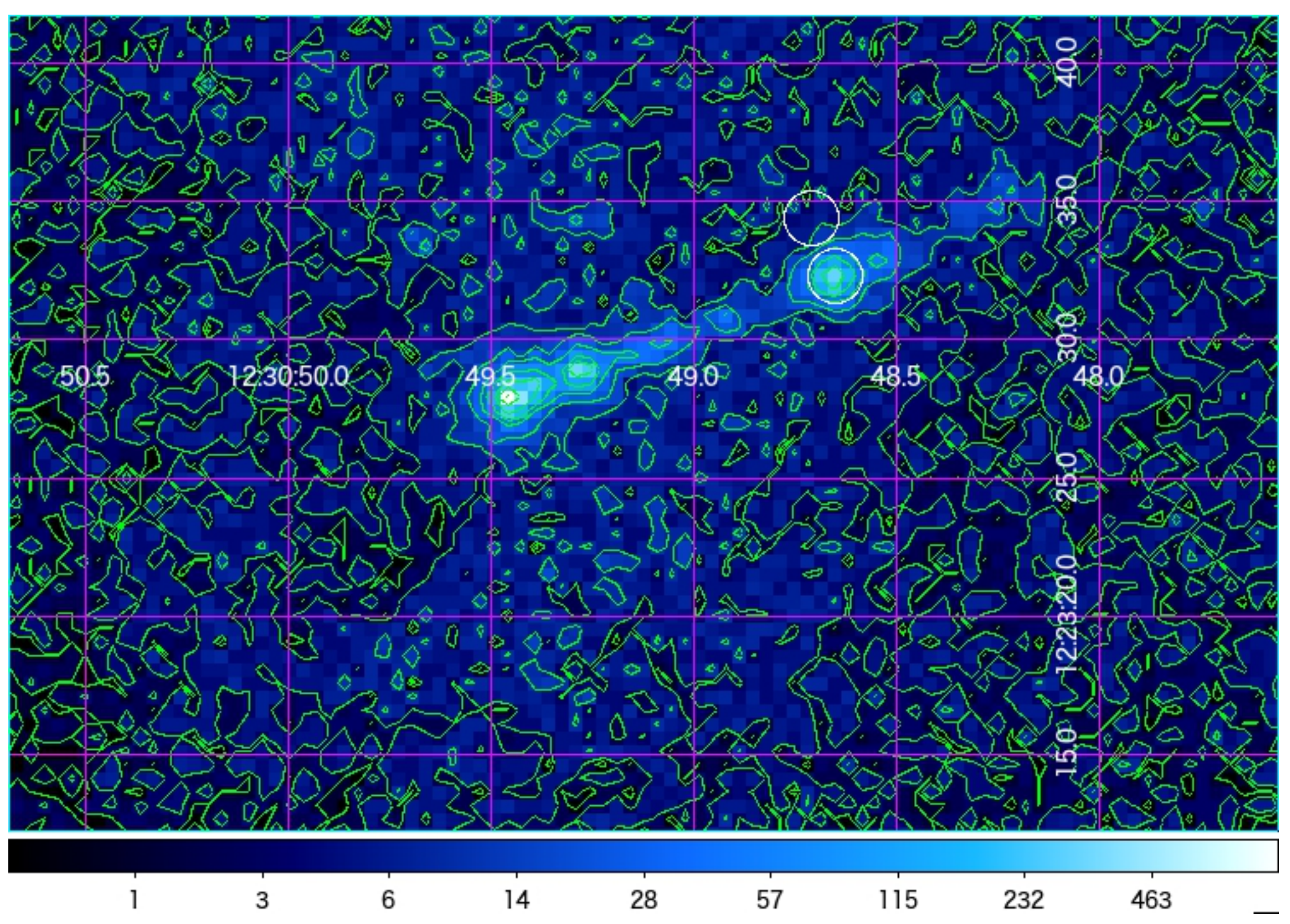

Figure 2 The smoothed image of M87 jet for obs ID 18232. A source region and background region for knot A in this analysis are shown by white circles.

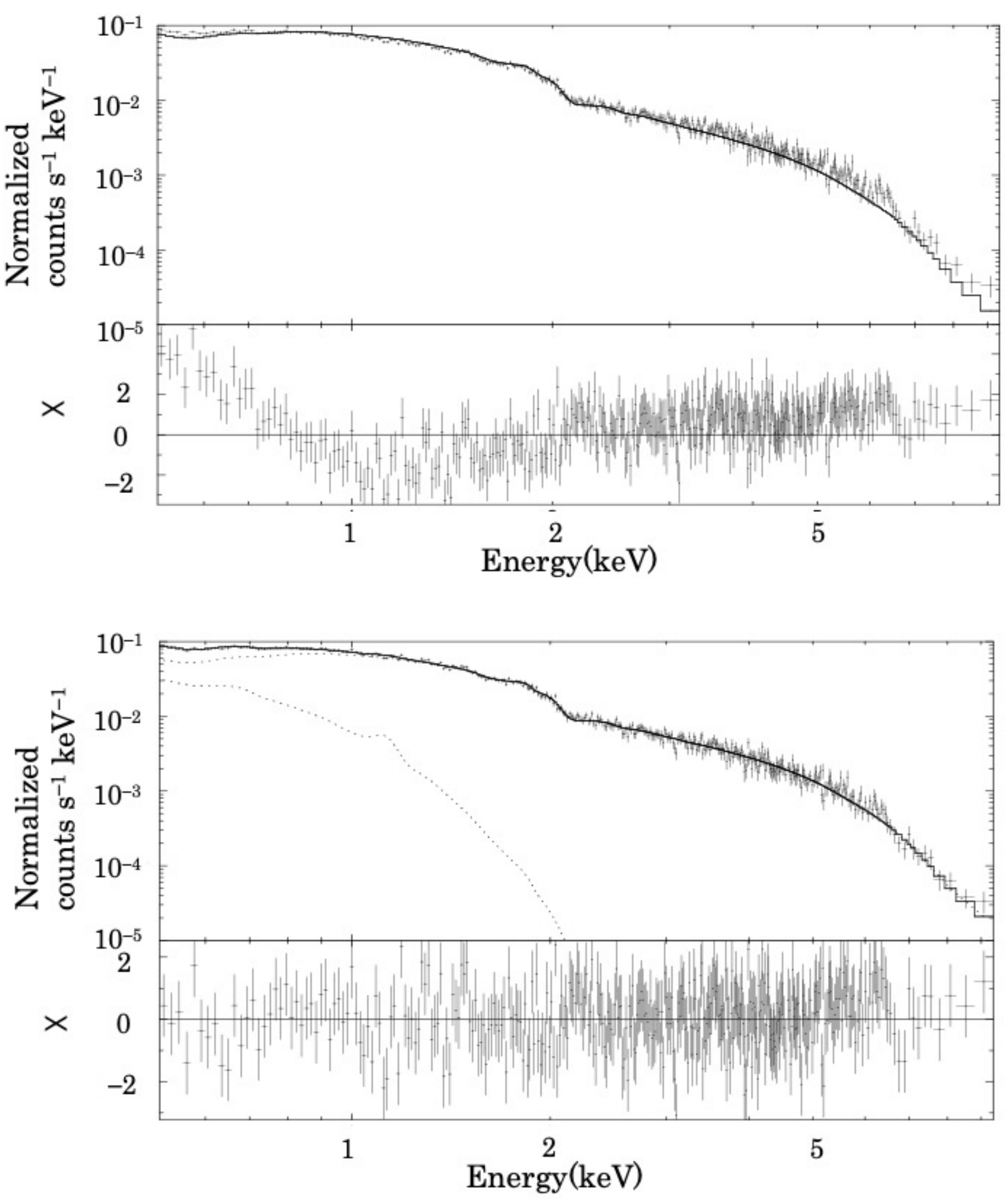

Figure 3 The fitted energy spectra with a power law model (top) and a combination model of a power law and an APEC (bottom) for all data of knot A. A column density of a photo absorption is fixed to the 21 cm radio observation value.